\documentclass[12pt,a4paper]{article}
\usepackage{amssymb}

\usepackage{epsfig}
\usepackage{graphicx}
\usepackage{amsmath}
\usepackage{float}

\newcommand\be{\begin{equation}}
\newcommand\ee{\end{equation}}
\newcommand\bea{\begin{eqnarray}}
\newcommand\eea{\end{eqnarray}}

\begin{document}

\title{Quantum Correlation Games}
\author{Azhar Iqbal and Stefan Weigert \\
%EndAName
HuMP -- Hull Mathematical Physics\\
Department of Mathematics, University of Hull, UK\\
\texttt{A.Iqbal@maths.hull.ac.uk}, \texttt{S.Weigert@hull.ac.uk}}
\date{March 2004 (vs2)}
\maketitle

\begin{abstract}
A new approach to play games quantum mechanically is proposed. We consider
two players who perform measurements in an EPR-type setting. The payoff
relations are defined as functions of \emph{correlations}, i.e. without
reference to classical or quantum mechanics. Classical bi-matrix games are
reproduced if the input states are classical and perfectly anti-correlated,
that is, for a \emph{classical} correlation game. However, for a \emph{%
quantum} correlation game, with an entangled singlet state as input,
qualitatively different solutions are obtained. For example, the Prisoners'
Dilemma acquires a Nash equilibrium if the players both apply a \emph{mixed}
strategy. It appears to be conceptually impossible to reproduce the
properties of quantum correlation games within the framework of classical
games.
\end{abstract}

PACS: 03.67.-a, 02.50.Le

\section{Introduction}

To process information has been conceived for a long time as a purely
mathematical task, independent of the carrier of information. However,
problems such as identifying a marked object in a database \cite{grover97}
or the factorization of large integer numbers \cite{shor94} are solved in a
highly efficient way if information is stored and processed quantum
mechanically. Hence, the theory of \emph{quantum information} came into
existence generalizing classical bits to \emph{qubits}: linear combinations
of classically incompatible states are possible, and they can be processed
simultaneously.

Game theory \cite{neumann+44}, a tool to take decisions in a rational way,
has been proposed as another promising candidate to benefit from a quantum
mechanical implementation \cite{eisert+99}. Based on their knowledge of the
circumstances, \emph{players} in a classical game select from a set of
possible \emph{moves} or \emph{actions} to maximize their \emph{payoffs}. In
its quantum version, unexpected moves may provide new \emph{solutions} to
the game; a strategy which includes quantum moves may outperform a classical
strategy \cite{meyer+99}. Opinions about the true \emph{quantum} character
of such games are divided, however. It has been argued that quantized games
are nothing but disguised classical games \cite{enk+02}. In other words, to
\emph{quantize} a game is claimed equivalent to replacing the original game
by a \emph{different} classical game.

In the present paper, we associate a quantum game with a classical game in a
way which addresses this criticism by imposing two constraints:

\begin{enumerate}
\item[(c1)]  The players choose their moves (or actions) from the \emph{same}
set in both the classical and the quantized game.

\item[(c2)]  The players agree on explicit expressions for their payoffs
which must \emph{not} be modified when switching between the classical and
the quantized version of the game.
\end{enumerate}

Games with these properties are expected to be immune against the criticism
raised above. In the new setting, the only `parameter' is the \emph{input
state} on which the players act, and its nature will determine the classical
or quantum character of the game. Our approach to quantum games, tailored to
satisfy both (c1) and (c2), is inspired by Bell's work \cite{bell64}: \emph{%
correlations} of measurement outcomes are essential. Effectively, we will
define payoff relations in terms of correlations - these payoffs will become
sensitive to the classical or quantum nature of the input allowing for
modified Nash equilibria.

Section $2$ introduces our notation of classical games. Then, games will be
set up in a way which resembles an EPR experiment. In Section $4$,
correlation games will be defined through payoffs depending explicitly on
correlations. If played on a classical input state, they reproduce classical
bi-matrix games. New advantageous strategies may emerge, however, if the
same payoff relations are used in the quantum mechanical setting, as shown
in Section $5$. Finally, we discuss achievements and limitations of our
approach.

\section{Matrix games and payoffs}

Consider a matrix game \cite{rasmusen89} for two players, called Alice and
Bob. A large set of identical objects are prepared in definite states, not
necessarily known to the players. Each object splits into two equivalent
`halves' handed over to Alice and Bob simultaneously. Let the players agree
beforehand on the following rules:

\begin{enumerate}
\item  Alice and Bob may either play the identity \emph{move} $I$ or perform
\emph{actions} $S_{A}$ and $S_{B}$, respectively. The moves $S_{A,B}$ (and $I
$) consist of unique actions such as flipping a coin (or not) and possibly
reading it.

\item  The players agree upon \emph{payoff relations} $P_{A,B}(p_{A},p_{B})$
which determine their awards as functions of their \emph{strategies}, that
is, the moves with probabilities $p_{A,B}$ assigned to them.

\item  The players fix their \emph{strategies} for repeated runs of the
game. In a \emph{mixed} strategy Alice plays the identity move $I$ with
probability $p_{A}$, say, while she plays $S_{A}$ with probability ${%
\overline{p}}_{A}=1-p_{A}$, and similarly for Bob. In a \emph{pure}
strategy, each player performs the same action in each run.

\item  Whenever the players receive their part of the system, they perform a
move consistent with their strategy.

\item  The players inform an arbiter about their actions taken in each
individual run. After a large number of runs, they are rewarded according to
the agreed payoff relations $P_{A,B}$. The existence of the arbiter is for
clarity only: alternatively, the players get together to decide on their
payoffs.
\end{enumerate}

These conventions are sufficient to play a classical game. As an example,
consider the class of symmetric bi-matrix games with payoff relations
\begin{eqnarray}
P_{A}(p_{A},p_{B}) &=&Kp_{A}p_{B}+Lp_{A}+Mp_{B}+N,  \notag \\
P_{B}(p_{A},p_{B}) &=&Kp_{A}p_{B}+Mp_{A}+Lp_{B}+N,  \label{payoffs}
\end{eqnarray}
where $K,L,M,$ and $N$ are real numbers. Being functions of two real
variables, $0\leq p_{A,B}\leq 1$, the payoff relations $P_{A,B}$ reflect
that each player may chose a strategy from a continuous one-parameter set.
The game is symmetric since
\begin{equation}
P_{A}(p_{A},p_{B})=P_{B}(p_{B},p_{A}).  \label{symmetry}
\end{equation}
Look at pure strategies with $p_{A,B}=0$ or $1$ in Eq. (\ref{payoffs}):
\begin{eqnarray}
P_{A}(1,1)=P_{B}(1,1) &=&r=K+L+M+N,  \notag \\
P_{A}(1,0)=P_{B}(0,1) &=&s=L+N,  \notag \\
P_{A}(0,1)=P_{B}(1,0) &=&t=M+N,  \notag \\
P_{A}(0,0)=P_{B}(0,0) &=&u=N,  \label{constants}
\end{eqnarray}
leading to the payoff \emph{matrix} for this game

\begin{equation}
\begin{array}{c}
\text{Alice}
\end{array}
\begin{array}{c}
I \\
S_{A}
\end{array}
\overset{\overset{
\begin{array}{c}
\text{Bob}
\end{array}
}{
\begin{array}{cc}
I & S_{B}
\end{array}
}}{\left(
\begin{array}{cc}
(r,r) & (s,t) \\
(t,s) & (u,u)
\end{array}
\right) .}  \label{matrix}
\end{equation}
In words: If both Alice and Bob play the identity $I$, they are paid $r$
units; Alice playing the identity $I$ and Bob playing $S_{B}$ pays $s$ and $%
t $ units to them, respectively; etc. Knowledge of the payoff matrix (\ref
{matrix}) and the probabilities $p_{A,B}$ is, in fact, equivalent to (\ref
{payoffs}) since the expected payoffs $P_{A,B}$ are obtained by averaging (%
\ref{matrix}) over many runs.

Let Alice and Bob act rationally: they will try to maximize their payoffs%
\footnote{%
The authors do not consider this the only possible definition of rationality.%
} by an appropriate strategy \cite{neumann+44}. If the entries of the matrix
(\ref{matrix}) satisfy $s<u<r<t$, the Prisoners' Dilemma \cite{rasmusen89}
arises: the players opt for strategies in which unilateral deviations are
disadvantageous; nevertheless, the resulting solution of the game, a Nash
equilibrium, does \emph{not} maximize their payoffs.

In view of the conditions (c1) and (c2) the form of the payoff relations $%
P_{A,B}$ in (\ref{payoffs}) seems to leave no room to introduce quantum
games which would differ from classical ones. In the following, we will
introduce payoff relations which \emph{are} sensitive to whether a game is
played on classical \emph{or} quantum objects. With classical input, they
will reproduce the classical game, and the conditions (c1) and (c2) will be
respected throughout.

\section{EPR-type setting of matrix games}

\emph{Correlation games} will be defined in a setting which is inspired by
EPR-type experiments \cite{peres93}. Alice and Bob are spatially separated,
and they share information about a Cartesian coordinate system with axes $%
\mathbf{e}_{x},\mathbf{e}_{y},\mathbf{e}_{z}$. The physical input used in a
correlation game is a large number of identical systems with zero angular
momentum, $\mathbf{J}=0$. Each system decomposes into a pair of objects
which carry perfectly anti-correlated angular momenta $\mathbf{J}_{A,B}$,
i.e. $\mathbf{J}_{A}+\mathbf{J}_{B}=0$.

In each run, Alice and Bob will measure the dichotomic variable $\mathbf{e}%
\cdot \mathbf{J}/|\mathbf{e}\cdot \mathbf{J}|$ of their halves along the
common $z$-axis ($\mathbf{e}\rightarrow \mathbf{e}_{z}$) or along specific
directions $\mathbf{e}\rightarrow \mathbf{e}_{A}$ and $\mathbf{e}\rightarrow
\mathbf{e}_{B}$ in two planes $\mathcal{P}_{A}$ and $\mathcal{P}_{B}$,
respectively, each containing the $z$-axis, as shown in Fig. $1$. The
vectors $\mathbf{e}_{A}$ and $\mathbf{e}_{B}$ are characterized by the
angles $\theta _{A}$ and $\theta _{B}$ which they enclose with the $z$-axis:
\begin{equation}
\mathbf{e}_{z}\cdot \mathbf{e}_{A,B}=\cos \theta _{A,B}\text{ },\qquad 0\leq
\theta _{A,B}\leq \pi .  \label{directions}
\end{equation}
In principle, Alice and Bob could be given the choice of both the directions
$\mathbf{e}_{A,B}$ \emph{and} the probabilities $p_{A,B}$. However, in
traditional matrix games each player has access to \emph{one} continuous
variable only, namely $p_{A,B}$. To remain within this framework, we impose
a relation between the probabilities $p_{A,B}\in \lbrack 0,1]$, and the
angles $\theta _{A,B}\in \lbrack 0,\pi ]$:
\begin{equation}
p_{A,B}=g(\theta _{A,B})\,.  \label{gfunction}
\end{equation}
The function $g$ maps the interval $[0,\pi ]$ to $[0,1]$, and it is
specified \emph{before} the game begins. This function is, in general, \emph{%
not} required to be invertible or continuous. Relation (\ref{gfunction})
says that Alice must play the identity with probability $p_{A}\equiv
g(\theta _{A})$ if she decides to select the direction $\mathbf{e}_{A}$ as
her alternative to $\mathbf{e}_{z}$; furthermore, she measures with
probability $\overline{p}_{A}=1-g(\theta _{A})$ along $\mathbf{e}_{A}$. For
an invertible function $g$, Alice can choose either a probability $p_{A}$ or
a direction $\theta _{A}$ and find the other variable from Eq. (\ref
{gfunction}). If the function $g$ is not invertible, some values of
probability are associated with more than one angle, and it is more natural
to have the players choose a direction first. For simplicity we will assume
the function $g$ to be invertible, if not specified otherwise.

According to her chosen strategy, Alice will measure the quantity $\mathbf{e}%
\cdot \mathbf{J}/|\mathbf{e}\cdot \mathbf{J}|$ with probability $p_{A}$
along the $z$-axis, and with probability $\overline{p}_{A}=1-p_{A}$ along
the direction $\mathbf{e}_{A}$. Similarly, Bob can play a mixed strategy,
measuring along the directions $\mathbf{e}_{z}$ or $\mathbf{e}_{B}$ with
probabilities $p_{B}$ and $\overline{p}_{B}$, respectively. Hence, Alice's
moves consists of either $S_{A}$ (rotating a Stern-Gerlach type apparatus
from $\mathbf{e}_{z}$ to $\mathbf{e}_{A}$, followed by a measurement) or of $%
I$ (a measurement along $\mathbf{e}_{z}$ with no previous rotation). Bob's
moves $I$ and $S_{B}$ are defined similarly. It is convenient to denote the
outcomes of measurements along the directions $\mathbf{e}_{A},\mathbf{e}_{B}$%
, and $\mathbf{e}_{z},$ by $a,b,$ and $c$, respectively.

%%%%%%%%%%%%%%%%%%%%%%%%%%%%%%%%%%%%%%%%%%%%%%%%%%%%%%%%%%%%%%%%%%
\begin{figure}[h]
\begin{center}
\includegraphics[width=.4\textwidth]{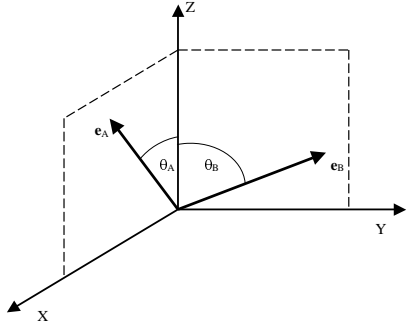}
\end{center}
\caption{The players' strategies consist of defining angles $\protect\theta%
_{A,B}$ which the directions $\mathbf{e}_{A,B} $ make with the
$z$-axis; for
simplicity, the planes $\mathcal{P}_{A,B}$ are chosen as the $xz$- and $yz$%
-plane, respectively.} \label{Fig1}
\end{figure}
%%%%%%%%%%%%%%%%%%%%%%%%%%%%%%%%%%%%%%%%%%%%%%%%%%%%%%%%%%%%%%%%%
%\FRAME{ftbpFU}{2.5149in}{1.8922in}{0pt}{\Qcb{The players' strategies consist
%of defining angles $\protect\theta _{A,B}$ which the directions $\mathbf{e}%
%_{A,B}$ make with the $z$-axis; for simplicity, the planes $\mathcal{P}_{A,B}
%$ are chosen as the $xz$- and $yz$-plane, respectively.}}{\Qlb{Fig1}}{%
%fig1.eps}{\special{language "Scientific Word";type
%"GRAPHIC";maintain-aspect-ratio TRUE;display "ICON";valid_file
%"F";width 2.5149in;height 1.8922in;depth 0pt;original-width
%5.1828in;original-height 3.8847in;cropleft "0";croptop
%"1";cropright "1";cropbottom "0";filename
%'Fig1.eps';file-properties "XNPEU";}}

After each run, the players inform the arbiter about the chosen directions
and the result of their measurements. After $N\rightarrow \infty $ runs of
the game, the arbiter possesses a list $\mathcal{L}$ indicating the
directions of the measurements selected by the players and the measured
values of the quantity $\mathbf{e}\cdot \mathbf{J}/|\mathbf{e}\cdot \mathbf{J%
}|$. The arbiter uses the list to determine the strategies played by Alice
and Bob by simply counting the number of times ($N_{A}$, say) that Alice
measured along $\mathbf{e}_{A}$, giving $p_{A}=\underset{N\rightarrow \infty
}{\lim }(N-N_{A})/N$, etc. Finally, the players are rewarded according to
the payoff relations (\ref{payoffs}).

\section{Correlation games}

We now develop a new perspective of matrix games in the EPR-type setting.
The basic idea is to define payoffs $P_{A,B}=P_{A,B}(\left\langle
ac\right\rangle ,\left\langle cb\right\rangle )$ which depend explicitly on
the \emph{correlations} of the actual measurements performed by Alice and
Bob. The arbiter will extract the numerical values of the correlations $%
\left\langle ac\right\rangle $ etc. from the list $\mathcal{L}$ in the usual
way. Consider, for example, all cases with Alice measuring along $\mathbf{e}%
_{A}$ and Bob along $\mathbf{e}_{z}$. If there are $N_{ac}$ such runs, the
correlation of the measurements is defined by
\begin{equation}
\left\langle ac\right\rangle =\lim_{N_{ac}\rightarrow \infty }\left(
\sum_{n=1}^{N_{ac}}\frac{a_{n}c_{n}}{N_{ac}}\right) \,,  \label{corraverage}
\end{equation}
where $a_{n}$ and $c_{n}$ take the values $\pm 1$ \cite{peres93}. The
correlations $\left\langle ab\right\rangle $ and $\left\langle
cb\right\rangle $ are defined similarly.

A symmetric bi-matrix correlation game is determined by a function $g$ in (%
\ref{gfunction}) and by the relations
\begin{eqnarray}
P_{A}(\left\langle ac\right\rangle ,\left\langle cb\right\rangle ) &=&K\text{
}G(\left\langle ac\right\rangle )G(\left\langle cb\right\rangle )+L\text{ }%
G(\left\langle ac\right\rangle )+M\text{ }G(\left\langle cb\right\rangle )+N,
\notag \\
P_{B}(\left\langle ac\right\rangle ,\left\langle cb\right\rangle ) &=&K\text{
}G(\left\langle ac\right\rangle )G(\left\langle cb\right\rangle )+M\text{ }%
G(\left\langle ac\right\rangle )+L\text{ }G(\left\langle cb\right\rangle )+N,
\label{re-expressed2}
\end{eqnarray}
where, in view of later developments, the function $G$ is taken to be
\begin{equation}
G(x)=g\left( \frac{\pi }{2}(1+x)\right) \,,\quad x\in \lbrack 0,1].
\label{Gfunction}
\end{equation}
As they stand, the payoff relations (\ref{re-expressed2}) do refer to
neither a classical nor a quantum mechanical input. Hence, condition (c2)
from above is satisfied: the payoff relations used in the classical and the
quantum version of the game are \emph{identical}, namely given by Eqs. (\ref
{re-expressed2}). Furthermore, Alice and Bob choose from the same set of
moves in both versions of the game: they select directions $\mathbf{e}_{A}$
and $\mathbf{e}_{B}$ (with probabilities $p_{A,B}$ associated with $\theta
_{A,B}$ via (\ref{gfunction})) so that condition (c1) is satisfied.
Nevertheless, the solutions of the correlation game (\ref{re-expressed2})
will depend on the input being either a classical or a quantum mechanical
anti-correlated state.

\subsection{Classical correlation games}

Alice and Bob play a \emph{classical correlation game} if they receive
classically anti-correlated pairs and use the payoff relations (\ref
{re-expressed2}). In this case, the payoffs turn into
\begin{equation}
P_{A,B}^{cl}=P_{A,B}(\left\langle ac\right\rangle _{cl},\left\langle
cb\right\rangle _{cl}),  \label{defineclcorrgame}
\end{equation}
where the correlations, characteristic for classically anti-correlated
systems \cite{peres93}, are given by
\begin{eqnarray}
\left\langle ac\right\rangle _{cl} &=&-1+2\theta _{A}/\pi ,  \notag \\
\left\langle cb\right\rangle _{cl} &=&-1+2\theta _{B}/\pi .  \label{correls}
\end{eqnarray}
%
%The angle $ \theta _{AB}=\arccos (\cos  \theta _{A}\cos  %\theta _{B})$ is defined by the directions %$\mathbf{e}_\alpha$ and $\mathbf{e}_\beta$.
Use now the definition of the function $G$ in (\ref{Gfunction}) and the link
(\ref{gfunction}) between probabilities $p_{A,B}$ and angles $\theta _{A,B}$
to obtain
\begin{eqnarray}
G(\langle ac\rangle ) &=&g(\theta _{A})=p_{A}, \\
G(\langle cb\rangle ) &=&g(\theta _{B})=p_{B}.  \label{G-p}
\end{eqnarray}
Hence, for classical input Eqs. (\ref{re-expressed2}) reproduce the payoffs
of a symmetric bi-matrix game (\ref{payoffs}),
\begin{eqnarray}
P_{A}^{cl}(p_{A},p_{B}) &=&Kp_{A}p_{B}+Lp_{A}+Mp_{B}+N,  \notag \\
P_{B}^{cl}(p_{A},p_{B}) &=&Kp_{A}p_{B}+Mp_{A}+Lp_{B}+N.  \label{clpayoffs}
\end{eqnarray}
The game-theoretic analysis of the classical correlation game is now
straightforward---for example, appropriate values of the parameters $%
(r,s,t,u)$ lead to the Prisoners' Dilemma, for \emph{any} invertible
function $g$.

\subsection{Quantum correlation games}

Imagine now that Alice and Bob receive quantum mechanical anti-correlated
singlet states
\begin{equation}
|\psi \rangle =\frac{1}{\sqrt{2}}\left( |+,-\rangle -|-,+\rangle \right) .
\label{singlet}
\end{equation}
They are said to play a \emph{quantum correlation game} if again they use
the payoff relations (\ref{re-expressed2}) which read in this case
\begin{equation}
P_{A,B}^{q}=P_{A,B}(\left\langle ac\right\rangle _{q},\left\langle
cb\right\rangle _{q}).  \label{defineqcorrgame}
\end{equation}
As before, Alice and Bob transmit the results of their measurements (on
their quantum halves) to the arbiter who, after a large number of runs,
determines the correlations $\left\langle ac\right\rangle _{q}$ and $%
\left\langle cb\right\rangle _{q}$ by the formula (\ref{corraverage})
\begin{eqnarray}
\left\langle ac\right\rangle _{q} &=&-\cos \theta _{A},  \notag \\
\left\langle cb\right\rangle _{q} &=&-\cos \theta _{B},  \label{qcorrels}
\end{eqnarray}
in contrast to (\ref{correls}).

The inverse of relation (\ref{gfunction}), then, links the probabilities and
correlations through
\begin{eqnarray}
\left\langle ac\right\rangle _{q} &=&-\cos \left( g^{-1}(p_{A})\right) ,
\notag \\
\left\langle cb\right\rangle _{q} &=&-\cos \left( g^{-1}(p_{B})\right) .
\label{re-expressed correl}
\end{eqnarray}
Plugging these expressions into the right-hand-side of (\ref{defineqcorrgame}%
), we obtain \emph{quantum} payoffs:
\begin{eqnarray}
P_{A}^{q}(p_{A},p_{B})
&=&KQ_{g}(p_{A})Q_{g}(p_{B})+LQ_{g}(p_{A})+MQ_{g}(p_{B})+N,  \notag \\
P_{B}^{q}(p_{A},p_{B})
&=&KQ_{g}(p_{A})Q_{g}(p_{B})+MQ_{g}(p_{B})+LQ_{g}(p_{A})+N.  \label{Qpayoffs}
\end{eqnarray}
where
\begin{equation}
Q_{g}(p_{A,B})=g\left( \frac{\pi }{2}\left( 1-\cos \left(
g^{-1}(p_{A,B})\right) \right) \right) \in \lbrack 0,1].  \label{expquantum}
\end{equation}
The payoffs $P_{A,B}^{q}$ turn out to be \emph{non-linear} functions of the
probabilities $p_{A,B}$ while the payoffs $P_{A,B}^{cl}$ of the classical
correlation game are \emph{bi-linear}. This modification has an impact on
the solutions of the game as shown in the following section.

\section{Nash equilibria of quantum correlation games}

What are the properties of the quantum payoffs $P_{A,B}^{q}$ compared to the
classical ones, $P_{A,B}^{cl}$? The standard approach to `solving games'
consists in studying Nash equilibria. For a bi-matrix game a pair of
strategies $(p_{A}^{\star },p_{B}^{\star })$ is a Nash equilibrium if each
players' payoff does not increase upon unilateral deviation from it,
\begin{eqnarray}
P_{A}(p_{A},p_{B}^{\star }) &\leq &P_{A}(p_{A}^{\star },p_{B}^{\star
})\,,\quad \text{for all }p_{A},  \notag \\
P_{B}(p_{A}^{\star },p_{B}) &\leq &P_{B}(p_{A}^{\star },p_{B}^{\star
})\,,\quad \text{for all }p_{B}.  \label{NE}
\end{eqnarray}
In the following, we will study the differences between classical and
quantum correlation games which are associated with two paradigmatic games:
the Prisoners' Dilemma (PD) and the Battle of Sexes (BoS).

The payoff matrix of the PD has been introduced in (\ref{matrix}). It will
be convenient to use the notation of game theory: $C\thicksim I$ corresponds
to Cooperation, while $D\thicksim S_{A,B}$ is the strategy of Defection. A
characteristic feature of this game is that the condition $s<u<r<t$
guarantees that the strategy $D$ dominates the strategy $C$ for both players
and that the unique equilibrium at $(D,D)$ is not Pareto optimal. An outcome
of a game is Pareto optimal if there is no other outcome that makes one or
more players better off and no player worse off. This can be seen in the
following way. The conditions (\ref{NE}) read explicitly
\begin{eqnarray}
0 &\leq &\left( Kp_{B}^{\star }+L\right) (p_{A}^{\star }-p_{A})\text{ }%
,\quad \text{for all }p_{A},  \notag \\
0 &\leq &\left( Kp_{A}^{\star }+L\right) (p_{B}^{\star }-p_{B})\text{ }%
,\quad \text{for all }p_{B}.  \label{NE1}
\end{eqnarray}
with $K$ and $L$ from (\ref{constants}). The inequalities have only one
solution
\begin{equation}
p_{A}^{\star }=p_{B}^{\star }=0\text{ },  \label{clpureNE}
\end{equation}
which corresponds to $(D,D)$, a pure strategy for both players. The PD is
said to have a pure Nash equilibrium.

The BoS is defined by the following payoff matrix:

\begin{equation}
\begin{array}{c}
\text{Alice}
\end{array}
\begin{array}{c}
I \\
S_{A}
\end{array}
\overset{\overset{
\begin{array}{c}
\text{Bob}
\end{array}
}{
\begin{array}{cc}
I & S_{B}
\end{array}
}}{\left(
\begin{array}{cc}
(\alpha ,\beta ) & (\gamma ,\gamma ) \\
(\gamma ,\gamma ) & (\beta ,\alpha )
\end{array}
\right) ,}
\end{equation}
where $I$ and $S_{A,B}$ are pure strategies and $\alpha >\beta >\gamma $.
Three Nash equilibria arise in the classical BoS, two of which are pure: $%
(I,I)$ and $(S_{A},S_{B})$. The third one is a mixed equilibrium where Alice
and Bob play $I$ with probabilities

\begin{equation}
p_{A}^{\star }=\frac{\alpha -\gamma }{\alpha +\beta -2\gamma }\text{ }%
,\qquad p_{B}^{\star }=\frac{\beta -\gamma }{\alpha +\beta -2\gamma }\text{ }%
.  \label{BoSMixedNash}
\end{equation}

For the \emph{quantum correlation game} associated with the generalized PD,
the conditions (\ref{NE}) turn into
\begin{eqnarray}
0 &\leq &\left( KQ_{g}(p_{B}^{\star })+L\right) \left( Q_{g}(p_{A}^{\star
})-Q_{g}(p_{A})\right) \,, \\
0 &\leq &\left( KQ_{g}(p_{A}^{\star })+L\right) \left( Q_{g}(p_{B}^{\star
})-Q_{g}(p_{B})\right) \,,  \label{QNEqualities}
\end{eqnarray}
where the range of $Q_{g}(p_{A,B})$ has been defined in (\ref{expquantum}).
Thus, the conditions for a Nash equilibrium of a quantum correlation game
are structurally similar to those of the classical game except for
non-linear dependence on the probabilities $p_{A,B}$. The only solutions of (%
\ref{QNEqualities}) therefore read
\begin{equation}
Q_{g}({p_{A}^{\star }})=Q_{g}({p_{B}^{\star }})=0\,,  \label{qpureNE}
\end{equation}
generating upon inversion a Nash equilibrium at
\begin{equation}
(p_{A}^{\star })_{q}=(p_{B}^{\star })_{q}=Q_{g}^{-1}(0)=g\left( \arccos
\left( 1-\frac{2}{\pi }g^{-1}(0)\right) \right) ,  \label{QNEofPD}
\end{equation}
where the transformed probabilities now come with a subscript $q$ indicating
the presence of quantum correlations. The location of this new equilibrium
depends on the actual choice of the function $g$, as is shown below.

Similar arguments apply to the pure Nash equilibria of the BoS game while
the mixed classical equilibrium (\ref{BoSMixedNash}) is transformed into

\begin{eqnarray}
(p_{A}^{\star })_{q} &=&Q_{g}^{-1}(p_{A}^{\star })=g\left( \arccos \left( 1-%
\frac{2}{\pi }g^{-1}(\frac{\alpha -\gamma }{\alpha +\beta -2\gamma })\right)
\right) ,  \notag \\
(p_{B}^{\star })_{q} &=&Q_{g}^{-1}(p_{B}^{\star })=g\left( \arccos \left( 1-%
\frac{2}{\pi }g^{-1}(\frac{\beta -\gamma }{\alpha +\beta -2\gamma })\right)
\right) .
\end{eqnarray}

When defining a quantum correlation game we need to specify a function $g$
which establishes the link between probabilities $p_{A,B}$ and angles $%
\theta _{A,B}$. We will study the properties of quantum correlation games
for $g$-functions of increasing complexity. In the simplest case, the
function $g$ is ($i$) continuous and invertible; next, we chose a function $%
g $ being ($ii$) invertible and discontinuous or ($iii$) non-invertible and
discontinuous. For simplicity, all examples are worked out for piecewise
linear $g$-functions. The generalization to smooth $g$-functions turns out
to be straightforward, and the results do not change qualitatively as long
as the $g$ function preserves its characteristic features.

\subsection*{($i$) Continuous and invertible g-functions}

\begin{figure}[h]
\begin{center}
\includegraphics[width=.4\textwidth]{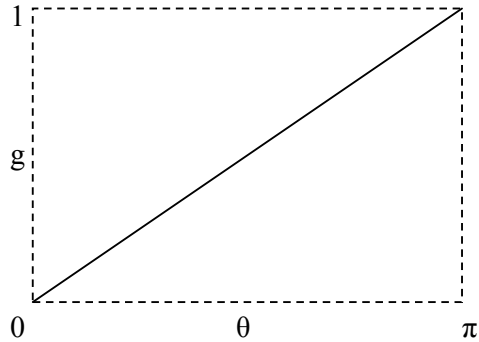}
\end{center}
\caption{The invertible and continuous $g$-function
$g_{1}(\protect\theta )=\protect\theta /\protect\pi$.}
\label{Fig2}
\end{figure}

Consider the function $g_{1}(\theta )=\theta /\pi $ defined for $\theta \in
\lbrack 0,\pi ]$. We have $g_{1}(0,\pi )=0$ or $1$, and the classical and
quantum correlations coincide at $\theta =0,\pi /2,$ and $\pi $. In view of (%
\ref{QNEofPD}) the function $g_{1}$ can have no effect on pure Nash
equilibria and the classical solution of PD is not modified in the quantum
game. Fig. \ref{Fig2} shows the function $g_{1}$.

%%%%%%%%%%%%%%%%%%%%%%%%%%%%%%%%%%%%%%%%%%%%%%%%%%%%%%%
%%%%%%%%%%%%%%%%%%%%%%%%%%%%%%%%%%%%%%%%%%%%%%%%%%%%%%%

%\FRAME{ftbpFU}{2.3021in}{1.6553in}{0pt}{\Qcb{A simple continuous and
%invertible $g$-function: $g_{1}(\protect\theta )=\protect\theta /\protect\pi
%$.}}{\Qlb{Fig2}}{fig2.eps}{\special{language "Scientific Word";type
%"GRAPHIC";maintain-aspect-ratio TRUE;display "ICON";valid_file "F";width
%2.3021in;height 1.6553in;depth 0pt;original-width 3.4359in;original-height
%2.4716in;cropleft "0";croptop "1";cropright "1";cropbottom "0";filename
%'Fig2.eps';file-properties "XNPEU";}}

However, solutions $p_{A,B}^{\star }\in (0,1)$ correspond to a mixed
classical equilibrium. It will be modified if $g(\pi /2)\neq p_{A,B}^{\star
} $ i.e. when the angle associated with $p_{A,B}^{\star }$ is different from
$\pi /2$. For example with the function $g_{1}(\theta )$ the probabilities
of the mixed equilibrium of the quantum correlation BoS are $(p_{A}^{\star
})_{q}=1-(1/\pi )\arccos \left\{ (\alpha -\gamma )/(\alpha +\beta -2\gamma
)\right\} $ and $(p_{B}^{\star })_{q}=1-(1/\pi )\arccos \left\{ (\beta
-\gamma )/(\alpha +\beta -2\gamma )\right\} $. A similar result holds for
the function $g_{2}(\theta )=1-\theta /\pi $.

\subsection*{($ii$) Invertible and discontinuous g-functions}

\begin{figure}[h]
\begin{center}
\includegraphics[width=.4\textwidth]{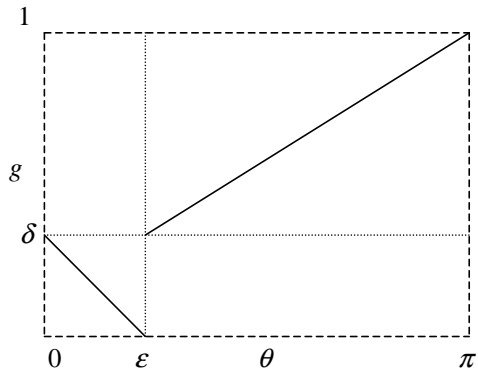}
\end{center}
\caption{Invertible and discontinuous $g$-function defined in Eq.
(\ref {InvDis1}).} \label{Fig3}
\end{figure}

For simplicity we consider invertible functions that are discontinuous at
one point only. Piecewise linear functions are typical examples. One such
function, shown in Fig. \ref{Fig3}, is
\begin{equation}
g_{3}(\theta )=\left\{
\begin{array}{rcl}
\delta (1-\theta /\epsilon ) & \text{if} & \theta \in \lbrack 0,\epsilon ]%
\text{ }, \\
\delta +(1-\delta )(\theta -\epsilon )/(\pi -\epsilon ) & \text{if} & \theta
\in (\epsilon ,\pi )\text{ }.
\end{array}
\right.   \label{InvDis1}
\end{equation}
The classical solution of PD $p_{A}^{\star }=p_{B}^{\star }=0$ disappears;
the new quantum solution is found at

\begin{equation}
(p_{A}^{\star })_{q}=(p_{B}^{\star })_{q}=\left\{
\begin{array}{rcl}
\delta +\frac{(1-\delta )}{(\pi -\epsilon )}\left\{ \arccos (1-2\epsilon
/\pi )-\epsilon \right\}  & \text{if} & \epsilon \in \lbrack 0,\frac{\pi }{2}%
]\text{ }, \\
\delta \left\{ 1-\frac{1}{\epsilon }\arccos (1-2\epsilon /\pi )\right\}  &
\text{if} & \epsilon \in (\frac{\pi }{2},\pi ]\text{ }.
\end{array}
\right.   \label{InvDis2}
\end{equation}
If, for example, $\delta =1/2$ and $\epsilon =\pi /4,$ we obtain a
mixed equilibrium at $(p_{A}^{\star })_{q}=(p_{B}^{\star
})_{q}=5/9.$ The appearance of a mixed equilibrium in a quantum
correlation PD game is an entirely non-classical feature.

%%%%%%%%%%%%%%%%%%%
%\FRAME{ftbpFU}{1.9268in}{1.4935in}{0pt}{\Qcb{The invertible and
%discontinuous $g$-function defined in Eq. (\ref{InvDis1}).}}{\Qlb{Fig3}}{%
%fig3.eps}{\special{language "Scientific Word";type
%"GRAPHIC";maintain-aspect-ratio TRUE;display "ICON";valid_file "F";width
%1.9268in;height 1.4935in;depth 0pt;original-width 2.9205in;original-height
%2.2632in;cropleft "0";croptop "1";cropright "1";cropbottom "0";filename
%'Fig3.eps';file-properties "XNPEU";}}

\begin{figure}[h]
\begin{center}
\includegraphics[width=.4\textwidth]{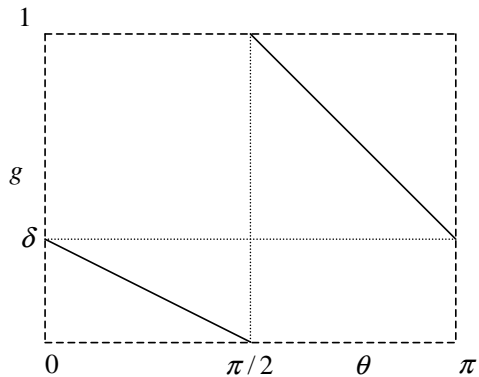}
\end{center}
\caption{Invertible and discontinuous $g$-function defined in Eq.
(\ref {InvDis3}).} \label{Fig4}
\end{figure}

The presence of a mixed equilibrium in the quantum correlation PD gives rise
to an interesting question: is there a Pareto-optimal solution of $(C,C)$ in
a quantum correlation PD with some invertible and discontinuous $g$%
-function? No such solution exists for invertible and continuous $g$%
-functions. Also, the $(C,C)$ equilibrium in PD cannot appear in a quantum
correlation game played with the function (\ref{InvDis1}): one has $%
g^{-1}(1)=\pi $ which can not be equal to $g^{-1}(0)$ when $g$ is
invertible. As a matter of fact, the solution $(C,C)$ for PD can be realized
in a quantum correlation PD if one considers $g$ from (\ref{InvDis2}) with $%
\epsilon =\pi /2$:

\begin{equation}
g_{4}(\theta )=\left\{
\begin{array}{rcl}
\delta (1-2\theta /\pi ) & \text{if} & \theta \in \lbrack 0,\frac{\pi }{2}]%
\text{ }, \\
1-2(1-\delta )(\theta -\pi /2)/\pi  & \text{if} & \theta \in (\frac{\pi }{2}%
,\pi ]\text{ },
\end{array}
\right.   \label{InvDis3}
\end{equation}
where $\delta \in (0,1)$, depicted in Fig. \ref{Fig4}.
%%%%%%%%%%%%%%%%
%\FRAME{ftbpFU}{1.9259in}{1.5307in}{0pt}{\Qcb{Invertible and discontinuous $g$%
%-function defined in Eq. (\ref{InvDis3}).}}{\Qlb{Fig4}}{fig4.eps}{\special%
%{language "Scientific Word";type "GRAPHIC";maintain-aspect-ratio
%TRUE;display "ICON";valid_file "F";width 1.9259in;height 1.5307in;depth
%0pt;original-width 2.8772in;original-height 2.2779in;cropleft "0";croptop
%"1";cropright "1";cropbottom "0";filename 'Fig4.eps';file-properties
%"XNPEU";}}

\begin{figure}[h]
\begin{center}
\includegraphics[width=.4\textwidth]{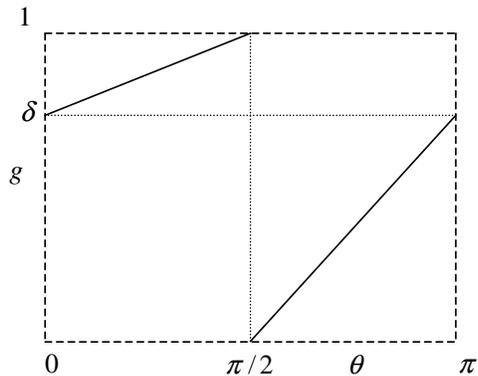}
\end{center}
\caption{Invertible and discontinuous $g$-function defined in Eq.
(\ref {InvDis4}).} \label{Fig5}
\end{figure}

This function satisfies $g^{-1}(0)=g^{-1}(1)=\pi /2$. Therefore, one has $%
\cos \left\{ g^{-1}(1)\right\} =1-2g^{-1}(0)/\pi $, which is the condition
for $(C,C)$ to be an equilibrium in PD. Cooperation $(C,C)$ will also be an
equilibrium in PD if the $g$-function is defined as

\begin{equation}
g_{5}(\theta )=\left\{
\begin{array}{rcl}
2(1-\delta )\theta /\pi +\delta  & \text{if} & \theta \in \lbrack 0,\frac{%
\pi }{2}]\text{ }, \\
2\delta (\theta -\pi /2)/\pi  & \text{if} & \theta \in (\frac{\pi }{2},\pi ]%
\text{ },
\end{array}
\right.   \label{InvDis4}
\end{equation}
where $\delta \in (0,1)$. Fig. \ref{Fig5} shows this function.
%%%%%%%%%%%
%\FRAME{ftbpFU}{1.8983in}{1.5022in}{0pt}{\Qcb{Invertible and discontinuous $g$%
%-function defined in Eq. (\ref{InvDis4}).}}{\Qlb{Fig5}}{fig5.eps}{\special%
%{language "Scientific Word";type "GRAPHIC";maintain-aspect-ratio
%TRUE;display "ICON";valid_file "F";width 1.8983in;height 1.5022in;depth
%0pt;original-width 2.8772in;original-height 2.2779in;cropleft "0";croptop
%"1";cropright "1";cropbottom "0";filename 'Fig5.eps';file-properties
%"XNPEU";}}

In both cases (\ref{InvDis3},\ref{InvDis4}), the $g$-function has a
discontinuity at $\theta =\pi /2$. With these functions both the pure and
mixed classical equilibria of BoS will also be susceptible to change. The
shifts in the pure equilibria in BoS will be similar to those of PD but the
mixed equilibrium of BoS will move depending on the location of $\delta $.

\begin{figure}[h]
\begin{center}
\includegraphics[width=.4\textwidth]{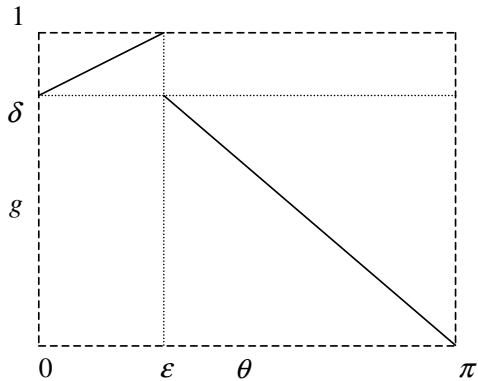}
\end{center}
\caption{Invertible and discontinuous $g$-function defined in Eq.
(\ref {InvDis5}).} \label{Fig6}
\end{figure}

Another example of an invertible and discontinuous function is given by

\begin{equation}
g_{6}(\theta )=\left\{
\begin{array}{rcl}
(1-\delta )\theta /\epsilon +\delta  & \text{if} & \theta \in \lbrack
0,\epsilon ]\text{ }, \\
\delta (\pi -\theta )/(\pi -\epsilon ) & \text{if} & \theta \in (\epsilon
,\pi ]\text{ },
\end{array}
\right.   \label{InvDis5}
\end{equation}
where $\delta \in (0,1)$ and $\epsilon \in (0,\pi )$ and it is drawn in Fig.
\ref{Fig6}.

%%%%%%%%%%%%%%%%%%%%%%%%%%%%%%%%%%
%\FRAME{ftbpFU}{1.8983in}{1.5316in}{0pt}{\Qcb{Invertible and discontinuous $g$%
%-function defined in Eq. (\ref{InvDis5}).}}{\Qlb{Fig6}}{fig6.eps}{\special%
%{language "Scientific Word";type "GRAPHIC";maintain-aspect-ratio
%TRUE;display "ICON";valid_file "F";width 1.8983in;height 1.5316in;depth
%0pt;original-width 2.834in;original-height 2.2779in;cropleft "0";croptop
%"1";cropright "1";cropbottom "0";filename 'Fig6.eps';file-properties
%"XNPEU";}}

\begin{figure}[h]
\begin{center}
\includegraphics[width=.4\textwidth]{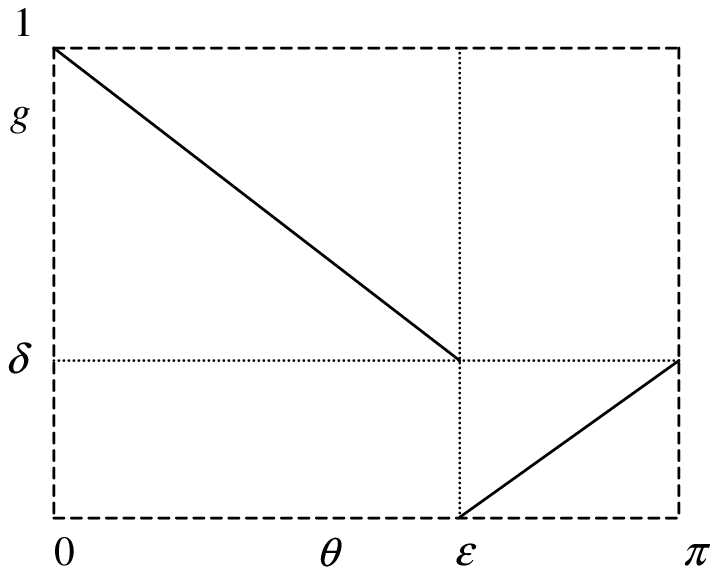}
\end{center}
\caption{Invertible and discontinuous $g$-function defined in Eq.
(\ref {InvDis6}).} \label{Fig7}
\end{figure}

With this function the pure classical equilibria $p_{A}^{\star
}=p_{B}^{\star }=0$ of PD as well as of BoS remain unaffected because these
equilibria require $\theta =\pi ,$ and the function is not discontinuous at $%
\pi $. One notices that if the angle corresponding to a classical
equilibrium is $0,\pi /2,$ or $\pi $, and there is no discontinuity at $\pi
/2$, then the quantum correlation game can not change that equilibrium. With
the function (\ref{InvDis5}) in both PD or BoS the pure equilibrium with $%
p_{A}^{\star }=p_{B}^{\star }=1$ corresponds to the angle $\theta =\epsilon $
where classical and quantum correlations are different (for $\epsilon \neq
\pi /2$). Consequently, the equilibrium $p_{A}^{\star }=p_{B}^{\star }=1$
will be shifted and the new equilibrium depends on the angle $\arccos
(1-2\epsilon /\pi )$. The mixed equilibrium of BoS will also be shifted by
the function (\ref{InvDis5}). Therefore, one of the pure equilibria and the
mixed equilibrium may shift if the $g$-function (\ref{InvDis5}) is chosen.
The following function
\begin{equation}
g_{7}(\theta )=\left\{
\begin{array}{rcl}
1-(1-\delta )\theta /\epsilon  & \text{if} & \theta \in \lbrack 0,\epsilon
]\,, \\
\delta (\theta -\epsilon )/(\pi -\epsilon ) & \text{if} & \theta \in
(\epsilon ,\pi ]\,,
\end{array}
\right.   \label{InvDis6}
\end{equation}
where $\delta \in (0,1)$ and $\epsilon \in (0,\pi ),$ cannot change the pure
equilibrium at $p_{A}^{\star }=p_{B}^{\star }=1$. However, it can affect the
equilibrium $p_{A}^{\star }=p_{B}^{\star }=1$, both in PD and BoS, and it
can shift the mixed equilibrium of BoS. Fig. \ref{Fig7} shows this function.

%%%%%%%%%%%%%%%%%%%%%%%%%%%%%%%
%\FRAME{ftbpFU}{1.8983in}{1.5316in}{0pt}{\Qcb{Invertible and discontinuous $g$%
%-function defined in Eq. (\ref{InvDis6}).}}{\Qlb{Fig7}}{fig7.eps}{\special%
%{language "Scientific Word";type "GRAPHIC";maintain-aspect-ratio
%TRUE;display "ICON";valid_file "F";width 1.8983in;height 1.5316in;depth
%0pt;original-width 2.834in;original-height 2.2779in;cropleft "0";croptop
%"1";cropright "1";cropbottom "0";filename 'Fig7.eps';file-properties
%"XNPEU";}}

\subsection*{($iii$) Non-invertible and discontinuous g-functions}

\begin{figure}[h]
\begin{center}
\includegraphics[width=.4\textwidth]{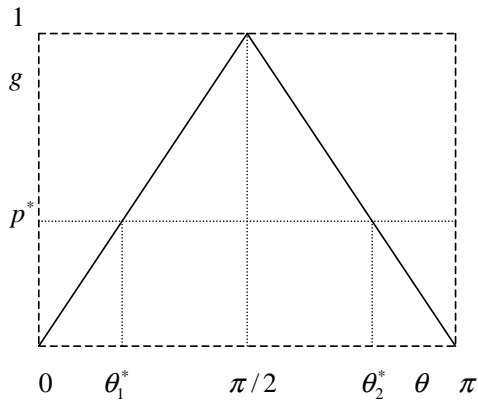}
\end{center}
\caption{Non-invertible and continuous $g$-function defined in Eq.
(\ref {NonInv1}).} \label{Fig8}
\end{figure}

A simple case of a continuous and non-invertible function (cf. Fig, \ref
{Fig8}) is given by
\begin{equation}
g_{8}(\theta )=\left\{
\begin{array}{rcl}
2\theta /\pi  & \text{if} & \theta \in \lbrack 0,\frac{\pi }{2}]\text{ ,} \\
1-2(\theta -\frac{\pi }{2})/\pi  & \text{if} & \theta \in (\frac{\pi }{2}%
,\pi ]\text{ .}
\end{array}
\right.   \label{NonInv1}
\end{equation}
Consider a classical pure equilibrium with $p_{A}^{\star }=p_{B}^{\star }=0$%
. Because $g^{-1}(0)=0$ or $\pi ,$ two equilibria with $g\left\{ \arccos
(\pm 1)\right\} $ are generated in the quantum correlation game, but these
coincide and turn out to be same as the classical ones. Similarly, the
function (\ref{NonInv1}) does not shift the pure classical equilibrium at $%
p_{A}^{\star }=p_{B}^{\star }=1$. However, if $p_{A,B}^{\star }\in (0,1)$
corresponds to a mixed equilibrium such that $g^{-1}(p^{\star })=\theta
_{1}^{\star },\theta _{2}^{\star }\neq \pi /2,$ then, in the quantum
correlation game, $p_{A,B}^{\star }$ will not only shift but also bifurcate.
The resulting values will differ from $p_{A,B}^{\star }$.

%%%%%%%%%%%%%%%%%%%%%%%%%%%%%%%%%%
%\FRAME{ftbpFU}{1.8706in}{1.5766in}{0pt}{\Qcb{Non-invertible and continuous $g
%$-function defined in Eq. (\ref{NonInv1}).}}{\Qlb{Fig8}}{fig8.eps}{\special%
%{language "Scientific Word";type "GRAPHIC";maintain-aspect-ratio
%TRUE;display "ICON";valid_file "F";width 1.8706in;height 1.5766in;depth
%0pt;original-width 2.834in;original-height 2.3886in;cropleft "0";croptop
%"1";cropright "1";cropbottom "0";filename 'Fig8.eps';file-properties
%"XNPEU";}}

Are the equilibria in a classical correlation game already susceptible to a
non-invertible and continuous $g$-function like (\ref{NonInv1})? When the
players receive the classical pairs of objects, the angles $\theta
_{1}^{\star },\theta _{2}^{\star }$ are mapped to themselves, resulting in
the same probability $p^{\star }$, obtained now using the non-invertible and
continuous $g$-function (\ref{NonInv1}). Therefore, in a classical
correlation game played with the function (\ref{NonInv1}) the bifurcation
observed in the quantum correlation game does not show up, in spite of the
fact that there are \emph{two} angles associated with one probability.

\section{Summary and Discussion}

\label{sec:Discussion}In this paper, we propose a new approach to introduce
a quantum mechanical version of bi-matrix games. One of our main objectives
has been to find a way to respect two constraints when `quantizing': on the
one hand, no new moves should emerge in the quantum game (c1) and, on the
other hand, the payoff relations should remain unchanged (c2). In this way,
we hope to circumvent objections which have been raised against existing
procedures to quantize games. New quantum moves or modified payoff relations
do not necessarily indicate a true quantum character of a game since their
emergence can be understood in terms of a modified \emph{classical} game.

\emph{Correlation games} are based on payoff relations which are sensitive
to whether the input is anti-correlated classically or quantum mechanically.
The players' allowed moves are fixed once and for all, and a setting
inspired by EPR-type experiments is used. Alice and Bob are both free to
select a direction in prescribed planes $\mathcal{P}_{A,B}$; subsequently
they individually measure, on their respective halves of the supplied
system, the value of a dichotomic variable either along the selected axis or
along the $z$-axis. When playing mixed strategies, they must use
probabilities which are related to the angles by a function $g$ which is
made public in the beginning. After many runs the arbiter establishes the
correlations between the measurement outcomes and rewards the players
according to fixed payoff relations $P_{A,B}$. The rewards depend only on
the numerical values of the \emph{correlations}---by definition, they do not
make reference to classical or quantum mechanics.

If the incoming states are classical, correlation games reproduce classical
bi-matrix games. The payoffs $P_{A,B}^{cl}$ and $P_{A,B}^{q}$ correspond to
one single game since both expressions emerge from the same payoff relation $%
P_{A,B}$. If the input consists of quantum mechanical singlet states,
however, the correlations turn quantum and the solutions of the correlation
game change. For example, in a generalized Prisoners' Dilemma a \emph{mixed}
Nash equilibrium can be found. This is due to an effective \emph{non-linear}
dependence of the payoff relations on the probabilities since the comparison
of Eqs. (\ref{clpayoffs}) and (\ref{Qpayoffs}) shows that `quantization'
leads to the substitution

\begin{equation}
p_{A,B}\rightarrow Q_{g}(p_{A,B})\,.  \label{quantize!}
\end{equation}
As the payoffs of traditional bi-matrix games are bi-linear in the
probabilities, it is difficult, if not impossible, to argue that the quantum
features of the quantum correlation game would arise from a disguised
classical game: there is no obvious method to let the payoffs of a classical
matrix game depend non-linearly on the strategies of the players.

Our analysis of the Prisoners' Dilemma and the Battle of Sexes as quantum
correlation games shows that, typically, both structure and location of
classical Nash equilibria are modified. The location of the quantum
equilibria depends sensitively on the properties of the function $g$ but,
apart from exceptional cases, the modifications are structurally stable. It
is \emph{not} possible to create any desired type of solution for a
bi-matrix game by a smart choice of the function $g$.

Finally, we would like to comment on the link between correlation games and
Bell's inequality. In spite of the similarity to an EPR-type experiment, it
is not obvious how to directly exploit Bell's inequality in correlations
games. Actually, its violation is \emph{not} crucial for the emergence of
the modifications in the quantum correlation game, as one can see from the
following argument. Consider a correlation game played on a mixture of
quantum mechanical anti-correlated \emph{product} states,
\begin{equation}
\hat{\rho}=\frac{1}{4\pi }\int_{\Omega }\,d\Omega \,|\mathbf{{e}_{\Omega
}^{+},{e}_{\Omega }^{-}\rangle \langle {e}_{\Omega }^{+},{e}_{\Omega
}^{-}|\,,}  \label{mixture}
\end{equation}
where the integration is over the unit sphere. The vectors $\mathbf{e}%
_{\Omega }^{\pm }$ are of unit length, and $|\mathbf{e}_{\Omega }^{\pm
}\rangle $ denote the eigenstates of the spin component $\mathbf{e}_{\Omega
}\cdot \hat{\mathbf{S}}$ with eigenvalues $\pm 1$, respectively. The
correlations in this entangled mixture are weaker than for the singlet state
$|\psi \rangle $
\begin{equation}
\left\langle ac\right\rangle _{\rho }=-\frac{1}{3}\cos \theta _{A}\,,\qquad %
\mbox{etc.}  \label{rhocorrels}
\end{equation}
The factor $1/3$ makes a violation of Bell's inequality impossible.
Nevertheless, a classical bi-matrix game is modified as before if $\hat{\rho}
$ is chosen as input state of the correlation game. To put this observation
differently: the payoffs introduced in Eq. (\ref{re-expressed2}) depend on
the two correlations $\langle ac\rangle $ and $\langle cb\rangle $ only, not
on the third one present in Bell's inequality, $\langle ab\rangle $.

An interesting development of the present approach consists of defining
payoffs of correlation games in such a way that they become sensitive to a
violation of Bell's inequality \cite{iqbal+03/1}. In this case, the
construction would assure that the game involves non-classical
probabilities, impossible to obtain by whatever classical game.

\end{document}